**DECISION SCIENCE INSTITUTE**
A Game-theoretic Model of the Consumer Behavior Under Pay-What-You-Want Pricing Strategy

Vahid Ashrafimoghari
Stevens Institute of Technology
vashraf1@stevens.edu

Jordan W. Suchow
Stevens Institute of Technology
jws@stevens.edu

**ABSTRACT**

In a digital age where companies face rapid changes in technology, consumer trends, and business environments, there is a critical need for continual revision of the business model in response to disruptive innovation. A pillar of innovation in business practices is the adoption of novel pricing schemes, such as Pay-What-You-Want (PWYW). In this paper, we employed game theory and behavioral economics to model consumers' behavior in response to a PWYW pricing strategy where there is an information asymmetry between the consumer and supplier. In an effort to minimize the information asymmetry, we incorporated the supplier's cost and the consumer's reference prices as two parameters that might influence the consumer's payment decision. Our model shows that consumers' behavior varies depending on the available information. As a result, when an external reference point is provided, the consumer tends to pay higher amounts to follow the social herd or respect her self-image. However, the external reference price can also decrease her demand when, in the interest of fairness, she forgoes the purchase because the amount she is willing to pay is less that what she recognizes to be an unrecoverable cost to the supplier.

**INTRODUCTION**

Over the past decades, technological shifts such as widespread availability of broadband internet, open source software, the sharing economy, big data, and the digitization of the products, on one hand, and influential management theories such as disruptive innovation and open innovation on the other hand, have encouraged corporations to rethink their business models. In fact, many studies show that disruptive innovation is a viable, or even the only, option to cope with rapid changes in a globalized economy (see Saki, 2016; Hopp et al., 2018). However, while revising pricing strategy should lie at the heart of any business's innovative practices, most companies stick to traditional pricing models such as cost-plus pricing and value-based pricing, primarily due to fear of revenue loss. Nevertheless, the technological shifts have transformed not only businesses, but consumers as well. Today, consumers have better real-time access to information and more options available than in the past. Hence, aligning with this shift in consumer behavior, businesses should explore new pricing strategies, e.g., participative pricing mechanisms that enable consumers to participate in the process of determining the price they pay for a product or service.

Pay-What-You-Want (PWYW) is a participative pricing strategy that empowers consumers to determine the price they pay and gives suppliers the opportunity to apply additional price discrimination, to position their products in the market, and to beat the competition (Kim et al., 2008; Fernandez and Nahata, 2009; Schmidt et al., 2015; Chao et al., 2015; Wagner, 2019; Chen & Wyer, 2020). As compared to other participative pricing mechanisms, such as Name-Your-Own-Price (NYOP), auctions, and bargaining (Hann & Terwiesch, 2003, Spann et al., 2004), in the PWYW mechanism the consumer has more control over pricing because there is no fixed price set by the supplier and so the buyer may offer any price (including zero) and the supplier cannot reject it (Dorn & Suessmair, 2017; Greiff and Egbert, 2018; Ariely et al., 2018). Though PWYW can be regarded as a particular form of voluntary contribution mechanism (see Natter & Kaufman, 2015; Lynn, 2017; Spann et al., 2018; Bluvestein and Raghubir, 2021), it differs from other voluntary contribution mechanisms such as tipping and donation because it applies to the core products and services of the supplier, not to its supplementary services, and the consumer pays for himself, not others.

Several studies have considered potential applications of PWYW pricing in industries such as hospitality (e.g. Matilla and Gao, 2016), food and drink (Riener and Traxler, 2012; Chawan, 2019), art (Bimberg et al, 2020; Golinski, 2020; Gross et al., 2021), and entertainment (Gneezy et al., 2010; Kim, et al 2014; Borg et al., 2020; Boonsiritomachai and Sud-on, 2021). Also, Chung (2017) refers to several real-world instances of organizations utilizing PWYW pricing, from restaurants and cafes such as Soul and Panera Bread to The Metropolitan Museum of Art.

Many studies highlighted the processes and drivers influencing PWYW pricing. Recent empirical studies looked at consumers' behavior in PWYW settings through different lenses like perceived product and service quality (Weisstein et al., 2019), perceived price fairness (Chung, 2017; Narwal and Nayak, 2020), altruism (Peschla et al., 2019) or social acceptability (Mills and Groeningen, 2021). Although, mixed results in some cases justifies further scrutiny (Stangl and Prayag, 2017; Gravert, 2017; Park et al., 2017; Spann et al, 2018). Moreover, the role that information asymmetries play in shaping payment decisions of consumers in PWYW settings is overlooked in most theoretical and empirical studies.

In addition to empirical research findings, theoretical considerations help to disentangle possible mechanisms at play in the PWYW price setting. Our goal, in this paper, is to propose a game-theoretic conceptual model representing how information asymmetries affect prices paid in a single consumer–single supplier interaction under the PWYW pricing mechanism. In particular, the model reveals that under certain conditions, PWYW pricing can be profitable and can be used, not only for short-term promotional purposes, but also as a practical pricing strategy in the long-term.

In the following section, we briefly summarize the recent literature on PWYW pricing and drivers which likely influence consumer payment decisions under this pricing scheme. In Section 3, relying on game theory, we first describe a linear model for a PWYW game and then extend to incorporate fairness and minimize information asymmetries that influence the effectiveness of PWYW pricing. The final section provides a conclusion and discusses likely future research directions.

**LITERATURE REVIEW**

While PWYW pricing has drawn the attention of scholars from Marketing, Operations Management and Economics disciplines (Gneezy et al., 2010; Gneezy et al., 2012; Kim et al., 2014; Kim et al., 2009; Schmidt et al., 2015; Chawan, 2019), the findings are sometimes ambiguous. Experimental studies have pointed out the importance of several motives that drive positive payments by consumers in the PWYW pricing setting, including product type (e.g. Weisstein et al., 2019), quality of service (e.g. Narwal and Nayak, 2020), fairness (e.g. Machado and Sinha, 2013), reference price (e.g. Roy et al., 2016), income (e.g. Kunter , 2015), loyalty (e.g. Marett et al. , 2012), social distance (Kim et al., 2014), self-image (e.g. Gneezy et al., 2012), price consciousness (e.g. Kim et al., 2009), inequity aversion (e.g. Schmidt et al., 2015), and usage (e.g. Marett et al, 2012), among other factors. Furthermore, some procedural conditions such as design variations (e.g. Christopher & Macado, 2019), the payment timing (e.g. Viglia et al., 2019), anonymous exchanges (e.g. Soule and Madrigal, 2015), time pressure (e.g. Sharma et al., 2020) or using textual cues (e.g. Kunter, 2015) affect the economic success of PWYW.

Reference points play a crucial role in making payment decisions. In general, there are two types of reference points: the External Reference Price (ERP) and the Internal Reference Price (IRP). The ERP can be defined as a price suggested by the supplier in the purchase environment based on the average price of the item in the market or what others paid (e.g. Soule & Madrigal, 2015; Gross et al., 2021). The IRP is defined as an internal price-evaluation scale that is rooted in the past payments for similar products or services, or personal preferences and values of the consumer (Gneezy et al., 2012; Khashay and Samahita, 2015; Narwal and Nayak, 2020). In a PWYW setting, both external and internal reference prices influence consumers' payment decisions significantly (Johnson and Cui, 2013; Nieto-García et al., 2017).

The effects on ERPs on payment decisions are mixed. On the one hand, the availability of an ERP can reduce the consumer's uncertainty in making payment decisions and lead to higher prices paid compared to when no reference price is presented (e.g. Weisstein et al., 2019). On the other hand, in some studies no significant relationship is found between external reference prices and the amount paid (Gneezy et al., 2012; Weisstein et al., 2019), and in some cases (e.g., when setting a minimum and maximum for the suggested price) ERPs may even have a negative impact on the price paid (Johnson & Cui, 2013; Roy et al., 2016; Gross et al., 2021).

In general, three variants of external reference price strategies are known in the literature: the minimum price set by the supplier, which consumers should not underbid (Sayman and Akcay, 2020); a maximum price, which consumers are not allowed to overbid (Ahunbay et al., 2020); and the provision of a short list of predefined prices, from which the consumers can choose the price level they are willing to pay (Christopher and Machado, 2019). Weisstein, Choi and Andersen (2019) found that when an ERP is absent, consumers who purchase hedonic products are more willing to contribute higher PWYW payments since they perceive the quality of the product higher. However, in presence of an ERP, consumers who purchase utilitarian products are primed to pay higher within the PWYW mechanism because they perceive it as a higher quality product. On the other hand, in a PWYW context, consumers often rely on their own internal reference price as a memory-based intuition helps them to estimate the value of the purchasing item and to determine the price they are willing to pay (Kim et al., 2009; Johnson and Cui, 2013). Through an empirical study, Roy et al. (2021) depicted that in the PWYW setting, customers' willingness to pay increases depending on their internal reference price, when they understand their payment is visible to the others.

When both internal and external reference points are available, the interplay between the two will influence the consumer's payment decision. Several empirical studies shed light on the mediating role of an external reference price, which prompts as an anchor in purchasing decisions, influencing consumers' willingness to pay and shifting their internal reference price towards a socially acceptable amount (Chandarshekaran & Grewel, 2006; Johnson and Cui, 2013; Roy et al., 2016 Weisstein, Choi and Andersen 2019). Kukla-Gryz and Zagorska (2017) conclude that in a PWYW context, where no external reference price is presented, consumers base their payment decisions on individual factors such as internal reference prices. Logically, seeking their own economic benefit, consumers use their decision power to minimize their loss and maximize their surplus. However, regarding loss aversion, as a cognitive bias that prompts consumers to prefer avoiding loss to a same-size gain (Kahnemann & Tversky, 1979), if a consumer perceives the external reference price as being above her internal reference point, she frames it as a "loss." By contrast, in the consumer's view, a price below the internal reference point is framed as a "gain". Consequently, in a PWYW setting, consumers have more of a tendency toward underpayment rather than overpayment, hence gaining instead of losing (Johnson & Cui, 2013).

Looking through the lens of distributive justice, Jang and Chu (2012) defined the distribution of the ratio of price-paid to willingness to pay as a measure of fairness and showed that there is a similarity between the distribution patterns of price-paid to that seen in dictator games. Krawczyk et al. (2015) show that consumers tend to match their payment with the mean price of the past contributions in order to follow social norms. Bettray, Sussemair and Dorn (2017) examined the relationship between the perceived price fairness and the payment in PWYW context, finding that consumers evaluate the fairness of the external reference price against their internal reference point.

**THE MODEL**

In this section, we describe the sequence of events in a consumer–supplier interaction within a PWYW context. To analyze the effect of the supplier's cost and likely losses on the price determined by the consumer, we attempt to isolate our model from other elements that might affect the behavior, such as positive reciprocity induced by switching to PWYW or a sunk costs bias. Referring to the literature review section, several researchers outline that fairness perceptions are important for prices paid by consumers. In our model we aimed to incorporate price fairness and minimize the impact of information asymmetries on prices paid under PWYW pricing.

The PWYW Game is a sequential game of incomplete information whereas a typical PWYW interaction happens between a single supplier that produces a product or service at unit cost $C$ and a single consumer who has an internal reference price, as a private information which represents the value of that product or service to her. The supplier's cost is recoverable, either through disposal or funding resources, or non-recoverable (sunk cost), but this information is unknown to the consumer. The consumer decides whether to buy the item or to forgo the purchase. If the consumer decides to buy the item, she chooses a price less than or equal to her IRP, pays it, and takes the product. In general, the resulting payoffs from the buying interaction would be $V - P$ for the consumer and $P - C$ for the supplier.

**General Assumptions**

For the proposed model, there are some considerations as follows. First, the supplier has no option to refuse a consumer's offer even if the proposed price is less than its production cost ($P$

< *C*), implying that the payoff of both the consumer and the supplier in this interaction is determined by the consumer. Second, the consumer's valuation of the product or service is determined based onto her internal reference price as "an average of the price most recently paid for a given good and the price usually paid for products of the same category", according to Kim et al. (2009) definition and is not changing in the short-term. Third, aiming to minimize the effect of product quality or experience on consumer's payment decision, we assume that the payment happens before consumption not after that. Forth, regarding loss aversion as underlying priority of consumer's economic decisions, we assumed that the payoff for consumers who choose to buy the item is always non-negative. Finally, for a purpose of simplicity we use one-consumer-one supplier setting rather than one-consumer-multiple-suppliers. As a result, the bystander effect will be removed from our model – as some studies report that this effect impacts the behavior in similar economic games (see Panchanathan et al., 2013).

**Description**

For simplicity, consumer's total utility from purchasing the good at price *P* is assumed linear according to the following:
$$U_B = V - P, \qquad (1)$$

where *V* is the item's perceived value, or alternatively, the consumer's internal reference price for the item.

Also, regarding the concept of opportunity cost, when a customer chooses to not buy, she saves as much as the price she would pay if she had chosen to buy, or *P*. So, her payoff of not buying the item would be:
$$U_{NB} = P \qquad (2)$$

On the other side, buying an item produced at cost *C* by the customer at price *P* yields a payoff, or alternatively, a margin as following:
$$M_B = P - C \qquad (3)$$

If the consumer decides to not buy the item, the supplier's payoff will be either zero or –*C*, depending on being type *R (with recoverable cost)* or typ*e S (with sunk cost),* respectively. Figure 1 depicts a schematic representation of the PWYG game.

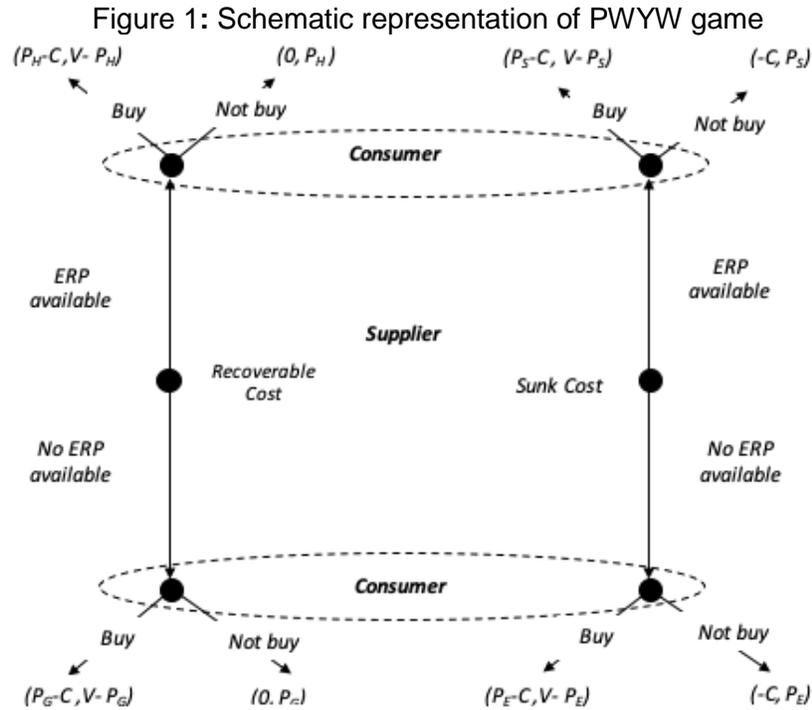

Figure 1: Schematic representation of PWYW game

As shown in Figure 1, depending on the supplier's type and its decision to whether decrease information asymmetry through providing an external reference price, optimal price paid by the consumer varies based on the following scenarios.

*Gain Seeking.* At the beginning, let's assume that the supplier provides no external reference price. So, if the consumer believes that the supplier is of type R, with recoverable cost, incentive for gain seeking would be the dominant driver of consumer's payment decision. The portion of free riders in this case will be probably highest amongst the four scenarios. Excluding a proportion of consumers who are free riders, let's assume each consumer has a unit demand and is fair-minded paying an amount, $P_G$ more than zero but less than her internal reference price, $V$. So, according to the Eq. (1), the payoff of buying for the consumer will be:

$$U_G = V - P_G \qquad (4)$$

At the same time, regarding Eq. (3) the payoff of buying the item by consumer at price $P_G$ for a supplier with cost $C$, or alternatively, the supplier's margin would be:

$$M_G = P_G - C \qquad (5)$$

Hence, if consumer decides to not purchasing the item, according to Eq. (2), her payoff will be:

$$U_{NG} = P_G \qquad (6)$$

and, given that the cost of supplier is recoverable, its payoff will be zero if the consumer decides not to buy the item.

*Herding.* Now, assume that the supplier provides information of an external reference price either in form of the average amount of the other consumer's payments or the average price of the similar item in the market. If consumer believes that the supplier is of type R, with recoverable cost, she follows the social norms by paying an amount, $P_H$, close to the external reference price, if it is less than her internal reference price for that item, $V$. In case that the external reference price is higher than consumer's internal reference price, she adjusts her payment only regarding her internal reference price. So, according to the Eq. (1), the payoff of buying for the consumer will be:

$$U_H = V - P_H \qquad (7)$$

At the same time, regarding Eq. (3) the payoff of buying the item by consumer at price $P_H$ for a supplier with cost $C$, or alternatively, the supplier's margin would be:

$$M_H = P_H - C \qquad (8)$$

Thus, if consumer decides to not purchasing the item, according to Eq. (2), her payoff will be:

$$U_{NH} = P_H \qquad (9)$$

Similar to the gain seeking scenario, since the cost of supplier is recoverable, its payoff will be zero, if consumer decides to not buying the item.

*Inequity Aversion.* In this scenario we assume the supplier suggests no external reference price for the item, but the consumer believes that being of type $S$, the supplier is unable to recover the product/service initial investment, through product disposal or funding resources. In this case, inequity aversion plays the most important role in consumer's payment decision because she concerns supplier's loss. Therefore, if the consumer chooses to buy, she pays $P_E$ close to her internal reference price and as a result, her payoff, regarding Eq. (1) would be:

$$U_E = V - P_E. \qquad (10)$$

At the same time, regarding Eq. (3) the payoff of buying the item by consumer at price $P_F$ for a supplier with cost $C$, or alternatively, the supplier's margin would be:

$$M_E = P_E - C \qquad (11)$$

Therefore, if consumer decides to not purchasing the item, according to Eq. (2), her payoff will be:

$$U_{NE} = P_E \qquad (12)$$

Since the supplier cannot recover its cost in this scenario, if consumer decides to not buying the item, its payoff would be as much as its production unit cost, $C$.

*Self-image.* In this scenario, we assume that the supplier provides an external reference price and like the Inequity Aversion scenario, the consumer believes that the supplier is unable to recover its production cost, if she chooses to not purchase. In this case, concerning her self-image, the consumer decides to pay $P_S$ close to the external reference price, only if it is less than her internal reference price. If the external reference price is higher than the consumer's internal reference price she decides to not buy to respect her self-image. So, the consumer's payoff, regarding Eq. (1) would be:

$$U_S = V - P_S \tag{13}$$

At the same time, regarding Eq. (3) the payoff of buying the item by consumer at price $P_F$ for a supplier with cost $C$, or alternatively, the supplier's margin would be:

$$M_S = P_S - C \tag{14}$$

Thus, if consumer decides to not purchasing the item, according to Eq. (2), her payoff will be:

$$U_{NS} = P_S \tag{15}$$

Since the supplier cannot recover its cost in this scenario, if consumer decides to not buying the item, its payoff would be as much as its production unit cost, $C$.

The above scenarios of consumer's response show that providing an external reference price, probably increases the amount paid by the consumer, tough it might decrease her demand. Now, if the supplier decides to reduce the information asymmetry further by revealing its cost to the consumer, the consumer will have additional reference points in hand to make her payment decision. To be specific, excluding a proportion of consumers who are free riders, let's assume that the consumer is fair-minded paying at least $C$ and therefore will not purchase the good if her internal reference price, $V$, is less than the supplier's cost, $C$. As a result, a fair-minded consumer splits the surplus $V - C$ out of reciprocity for the supplier having chosen a PWYW scheme (Schmidt et al., 2014). Let $\lambda$ be the proportion of surplus shared with supplier, $0 < \lambda \leq 1$. This parameter represents the strength of social preferences in the economy, and can also be interpreted as an exogenous social norm — typically assumed to be 0.5 in an equal sharing rule, but in a richer and more generous economy the norm may be to give more and vice versa (see, for example, Mrkava et al. (2020) who find cross-cultural variations in reciprocity).

In a one-supplier-one consumer interaction, the fair-minded consumer's PWYW payment is therefore defined to be:

$$P_{fair} = Ci\lambda(V - C) \tag{16}$$

Substituting (13) into the Eq. (1) then gives the fair-minded consumer's PWYW utility when the supplier's cost is known:

$$U_{fair} = (1 - \lambda)(V - C) \tag{17}$$

In general, let's assume that consumer's utility from purchasing the product is determined by two factors: an internal surplus (the difference between consumer's internal reference price, $V$, and the price she pays to the supplier, $P$) and utility loss from unfairness (the difference between the price $P$ and the fair price). We build our model based on the (Fehr and Schmidt, 1999) parametrized inequity aversion model through adding the possibility of consumer concerns supplier's loss:

$$U(P_R, P, C) = (P_R - P) - \alpha \max\{(P - C) - (P_R - P), 0\} - \beta \max\{(P_R - P) - (P - C), 0\} - \gamma \max\{C - P, 0\} \tag{18}$$

where $P_R$ represents consumer's reference price (internal or external), $P$ denotes price paid by the consumer, $C$ is the unit production cost for supplier and, $\alpha$ and $\beta$ are envy and altruism parameters that determine his or her attitude towards the disadvantageous and advantageous inequality respectively and, parameter $\gamma \geq 0$ determines consumer's sensitivity to the supplier's losses. That is, when a price is decreased by one dollar, the consumer's psychological penalty decreases by $\alpha$ if her surplus is lower than supplier's, and increases by $\beta$ if her surplus is higher than the supplier's, but the supplier's payoff is still positive. $\alpha$ will be further referred to as an "envy factor" and $\beta$ will be referred to as a "altruism factor". The model assumes $\alpha \geq \beta$ and $0 \leq \beta < 1$ (see (Fehr and Schmidt, 1999) for justification of these assumptions). If a supplier suffers losses, a consumer experiences a penalty of $\beta + \gamma$ for every dollar decline of the price. $\gamma = 0$ means that the consumer evaluates supplier's gains and losses equally, and our model is reduced to a classic FS model.

Let $P_f$ be the "fair price" – a price that results in an equitable split of the surplus created between the seller and the consumer: $P_f = 0.5(P_R + C)$. When $P = P_f$, the consumer does not suffer any unfairness-related utility loss. Let us further assume that the price is limited to be $P \geq 0$.

Then the optimal price for consumer would be:

$$P = \begin{cases} 0, & \text{if } \beta < 0.5 \text{ and } \beta + \gamma < 1; \\ C, & \text{if } \beta < 0.5 \text{ and } \beta + \gamma > 1; \\ P_f, & \text{if } \beta > 0.5. \end{cases} \quad (19)$$

If $\beta = 0.5$, the consumer is indifferent between any price within the interval $[C, P_f]$; if $\beta < 0.5$ and $\beta + \gamma = 1$, the consumer is indifferent between any price in the interval $[0, P_f]$. Note, that if $\gamma = 0$, like in the original Fehr and Schmidt model, $\beta < 0.5$ implies $\beta + \gamma < 1$, and the option $P = C$ can never be selected.
.

**DISCUSSION AND CONCLUSIONS**

Disruptive innovation in theory and recent technological changes in practice push businesses to adopt innovative pricing strategies like Pay-What-You-Want. This paper discusses different scenarios of consumer's response to PWYW price setting regarding the most important drivers that influence consumers' payment decisions. Therefore, we proposed a linear model of interaction between one supplier and one consumer under a PWYW price setting. We incorporated two parameters that might affect the price paid by the consumers under PWYW: the supplier's cost and the consumer's reference prices. Literature review shows that presence of reference prices is one of the most important drivers affecting the consumer's payment decisions. Also, if the supplier's unit production cost is known to the consumer, information asymmetry is reduced. Moreover, some studies argue that empathy can induce altruistic behavior, and conclude that empathy concerns increase sharing in interactions similar to dictator game (Von Bieberstein et al., 2021). Assuming a consumer does indeed empathize with the supplier's loss aversion, some of the buyers who would otherwise make the payment decision based on their own internal reference price may raise their prices if they learn about the supplier's costs, ensuring that the supplier will not go out of business after a while. This aspect has not been mentioned in previous interpretations of empirical findings on PWYW pricing and can complement existing models.

Our proposed conceptual model resembles the dictator game. In a typical dictator game (Bolton et al., 1998; Bardsley, 2008), there are two kind of players, dictator and receiver. The dictator has full power on deciding how to allocate a given endowment between herself and the other player. The receiver must accept the dictator's offer, even if the amount offered is zero. Game theory predicts that a rational dictator seeking to maximize her gain, should keep the whole amount for herself and transfer nothing to the receiver. However, most dictators do not maximize their payoffs transfer positive amounts to recipients instead (see Camerer, 2003; Oberholzer-Gee & Eichenberger, 2008; Korenack et al., 2013). The average transfer is reported to be 20 percent of the endowment (Camerer, 2003; Ploner & Regner, 2013). It shows that, even in a laboratory setting, the dictators have concerns beyond pure gain seeking.

However, there are some differences between our PWYW game and the standard dictator game. First, PWYW game outcomes are more realistic since you can hardly find a real world interaction happening based on dictator game, but many interactions in reality can be modeled using PWYW game. The second difference is that in a typical dictator game the payoffs are restricted to be non-negative, but in a PWYW game the supplier's payoff from an interaction with a consumer could be negative where consumer chooses to pay a price less than the supplier's production cost. The third difference is that if the price that a consumer is willing to pay is below the supplier's cost, the consumer might refuse to buy the good because of self-image concerns (see Gneezy et al., 2012; Kahsay & Samahita, 2015).

For future research, it is suggested to study supplier's profitability regarding the proposed model. Also, the model needs to be tested empirically to confirm the assumptions about the effect of reference prices and the supplier's cost on consumer's payment decisions.

**REFERENCES**

[1] Ahunbay, M. Ş., Lucier, B., & Vetta, A. (2020, September). Two-buyer sequential multiunit auctions with no overbidding. *In International Symposium on Algorithmic Game Theory* (pp. 3-16). Springer, Cham.

[2] Alford, B. L., & Biswas, A. (2002). The effects of discount level, price consciousness and sale proneness on consumers' price perception and behavioral intention. *Journal of Business Research*, 55(9), 775–783.

[3] Ariely, D., Gneezy, U. and Haruvy, E. (2018), Social Norms and the Price of Zero. *Journal of Consumer Psychology*, 28: 180–191.

[4] Bearden, W. O., Carlson, J. P., & Hardesty, D. M. (2003). Using invoice price information to frame advertised offers. *Journal of Business Research*, 56(5), 355–366.

[5] Bardsley, N. (2008). Dictator game giving: altruism or artefact?. Experimental Economics, 11(2), 122-133.

[6] Bluvstein Netter, S., & Raghubir, P. (2021). Tip to Show Off: Impression Management Motivations Increase Consumers' Generosity. *Journal of the Association for Consumer Research*, 6(1), 120–129.

[7] Boonsiritomachai, W., & Sud-on, P. (2021). PWYW Entrance fees: a visitor's perspective on a prominent art museum in Thailand. *Tourism Recreation Research*, 1–15.

[8] Dodds, W.B., Monroe, K.B., Grewal, D., 1991. "Effects of price, brand, and store information on consumers' product evaluations". Journal of Marketing Research. 28 (3), 307 –319.

[9] Dorn, T. and Suessmair, A. (2017) Determinants in Pay-What-You-Want Pricing Decisions—A Cross-Country Study. *American Journal of Industrial and Business Management*, 7, 115-142.

[10] Chandran, S., & Morwitz, V. G. (2005). Effects of participative pricing on consumers' cognitions and actions: A goal theoretic perspective. *Journal of Consumer Research*, 32(2), 249-259.

[11] Chawan, V. (2019). A pay-what-you-want pricing model for restaurants. *International Journal of Services and Operations Management*, 32(4), 431-449.

[12] Chen, Y., & Wyer Jr, R. S. (2020). The effects of endorsers' facial expressions on status perceptions and purchase intentions. *International Journal of Research in Marketing*, 37(2), 371-385.

[13] Chung, J. Y. (2017). Price fairness and PWYW (pay what you want): a behavioral economics perspective. *Journal of Revenue and Pricing Management*, 16(1), 40–55.

[14] Christopher, R. M., & Machado, F. S. (2019). Consumer response to design variations in pay-what-you-want pricing. *Journal of the Academy of Marketing Science*, 47(5), 879–898.

[15] Gerpott, T. (2017). Pay-what-you-want pricing: An integrative review of the empirical research literature. *Management Science Letters*, 7(1), 35–62.

[16] Gneezy, A., Gneezy, U., Nelson, L. D., & Brown, A. (2010). Shared social responsibility: A field experiment in pay-what-you-want pricing and charitable giving. *Science*, 329(5989), 325-327.

[17] Gneezy, A., Gneezy, U., Riener, G., & Nelson, L. D. (2012). Pay-what-you-want, identity, and self-signaling in markets. *Proceedings of the National Academy of Sciences*, 109(19), 7236–7240.

[18] Golinski, K. L. (2020). Voluntary Payments in Music: The Future of Creative Economies?. University of California, San Diego.

[19] Gravert, C. (2017). Pride and patronage-pay-what-you-want pricing at a charitable bookstore. *Journal of Behavioral and Experimental Economics*, 67, 1–7.

[20] Greiff, M., & Egbert, H. (2018). A review of the empirical evidence on PWYW pricing. Economic and Business Review, 20(2), 169–193.

[21] Hann, I. H., & Terwiesch, C. (2003). Measuring the frictional costs of online transactions: The case of a name-your-own-price channel. Management Science, 49(11), 1563–1579.

[22] Hopp, C., Antons, D., Kaminski, J., & Oliver Salge, T. (2018). Disruptive innovation: Conceptual foundations, empirical evidence, and research opportunities in the digital age. Journal of Product Innovation Management, 35(3), 446-457.

[23] Jang, H., & Chu, W. (2012). Are consumers acting fairly toward companies? An examination of pay-what-you-want pricing. Journal of Macromarketing, 32(4), 348–360.

[24] Johnson, J. W., & Cui, A. P. (2013). To influence or not to influence: External reference price strategies in pay-what-you-want pricing. Journal of Business Research, 66(2), 275–281.

[25] Kahneman, D., & Tversky, A. (1979). Interpretation of intuitive probability: a reply to Jonathan Cohen. Cognition, 7(4), 409-411.

[26] Kahsay, G. A., & Samahita, M. (2015). Pay-What-You-Want pricing schemes: A self-image perspective. *Journal of Behavioral and Experimental Finance*, 7, 17–28.

[27] Kim, J. Y., Kaufmann, K., & Stegemann, M. (2014). The impact of buyer–seller relationships and reference prices on the effectiveness of the pay what you want pricing mechanism. *Marketing Letters*, 25(4), 409–423.

[28] Kim, J.Y., Natter, M. and Spann, M. (2009) Pay What You Want: A New Participative Pricing Mechanism. *Journal of Marketing*, 73, 44–58.

[29] Kim, J.Y., Natter, M. and Spann, M. (2010) Kish: Where Consumers Pay as They Wish. *Review of Marketing Science*, 8, 1–12.

[30] Kim, J.Y., Natter, M. and Spann, M. (2014) Sampling, Discounts or Pay-What-You-Want: Two Field Experiments. *International Journal of Research in Marketing*, 31, 327–334.

[31] Korenok, O., Millner, E. L., & Razzolini, L. (2014). Taking, giving, and impure altruism in dictator games. *Experimental Economics*, *17*(3), 488-500.

[32] Krämer, F., Schmidt, K. M., Spann, M., & Stich, L. (2017). Delegating pricing power to customers: Pay what you want or name your own price? *Journal of Economic Behavior & Organization*, 136, 125-140.

[33] Krawczyk, M., Kukla-Gryz, A., & Tyrowicz, J. (2015). Pushed by the crowd or pulled by the leaders? Peer effects in Pay-What-You-Want (No. 2015-25). Faculty of Economic Sciences, University of Warsaw. [34] Kunter, M. (2015). Exploring the pay-what-you-want payment motivation. *Journal of Business Research*, 68(11), 2347–2357.

[35] Lynn, M. (2017). Should US restaurants abandon tipping? A review of the issues and evidence. *Psychosociological Issues in Human Resource Management*, 5(1), 120–159.

[36] Machado, F., & Sinha, R. K. (2013). The viability of pay what you want pricing. Management Science Working Paper.

[37] Marett, K., Pearson, R., & Moore, R. S. (2012). Pay what you want: An exploratory study of social exchange and buyer-determined prices of iProducts. *Communications of the Association for Information Systems*, 30(1), 10.

[38] Mrkva, K., Johnson, E. J., Gächter, S., & Herrmann, A. (2020). Moderating loss aversion: Loss aversion has moderators, but reports of its death are greatly exaggerated. *Journal of Consumer Psychology, 30(3)*, 407-428.

[39] Nieto-García, M., Muñoz-Gallego, P. A., & González-Benito, Ó. (2017). Tourists' willingness to pay for an accommodation: The effect of eWOM and internal reference price. *International Journal of Hospitality Management*, 62, 67–77.

[40] Narwal, P., & Nayak, J. K. (2020). Investigating relative impact of reference prices on customers' price evaluation in absence of posted prices: a case of Pay-What-You-Want (PWYW) pricing. *Journal of Revenue and Pricing Management*, 19(4), 234–247.

[41] Natter, M., & Kaufmann, K. (2015). Voluntary market payments: Underlying motives, success drivers and success potentials. *Journal of Behavioral and Experimental Economics*, 57, 149–157.

[42] Oberholzer-Gee, F., & Eichenberger, R. (2008). Fairness in extended dictator game experiments. *The BE Journal of Economic Analysis & Policy*, *8*(1).

[43] Panchanathan, K., Frankenhuis, W. E., & Silk, J. B. (2013). The bystander effect in an N-person dictator game. *Organizational Behavior and Human Decision Processes*, *120*(2), 285–297.

[44] Park, S., Nam, S., & Lee, J. (2017). Charitable giving, suggestion, and learning from others: Pay-What-You-Want experiments at a coffee shop. *Journal of Behavioral and Experimental Economics*, 66, 16–22.

[45] Ploner, M., & Regner, T. (2013). Self-image and moral balancing: An experimental analysis. *Journal of Economic Behavior & Organization*, *93*, 374–383.

[46] Rossi, P.H., 1979. 14. Vignette analysis: uncovering the normative structure of complex judgments. In R.K. Merton, J.S. Coleman and P.H.Rossi (Eds.), *Qualitative and quantitative social research: Papers in honor of Paul F. Lazarsfeld* (pp.176-186), The Free Press.

[47] Roy, R., Rabbanee, F. K., & Sharma, P. (2016). Antecedents, outcomes, and mediating role of internal reference prices in pay-what-you-want (PWYW) pricing. *Marketing Intelligence & Planning*, *34*(1), 117-136.

[48] Saki, Z. (2016). Disruptive innovations in manufacturing–an alternative for re-shoring strategy. *Journal of Textile and Apparel, Technology and Management*, *10*(2).

[49] Sayman, S., & Akçay, Y. (2020). A transaction utility approach for bidding in second-price auctions. *Journal of Interactive Marketing*, 49, 86–93.

[50] Schmidt, K. M., Spann, M., & Zeithammer, R. (2015). Pay what you want as a marketing strategy in monopolistic and competitive markets. *Management Science*, 61(6), 1217–1236.

[51] Spann, M., Skiera, B., & Schäfers, B. (2004). Measuring individual frictional costs and willingness-to-pay via name-your-own-price mechanisms. *Journal of Interactive Marketing*, 18(4), 22–36.

[52] Spann, M., Zeithammer, R., Bertini, M., Haruvy, E., Jap, S. D., Koenigsberg, O & Thomas, M. (2018). Beyond posted prices: the past, present, and future of participative pricing mechanisms. *Customer Needs and Solutions*, 5(1), 121–136.

[53] Stangl, B., & Prayag, G. (2017). Collaborative destination marketing and PWYW. *Annals of Tourism Research*, 70, 103–104.

[54] Thaler, R. H., Tversky, A., & Kahnemann, D. A. Schwartz (1997), The effect of myopic and loss aversion on risk taking: an experimental test. *The Quarterly Journal of Economics*, May, 647–660.

[55] Von Bieberstein, F., Essl, A., & Friedrich, K. (2021). Empathy: A clue for prosocialty and driver of indirect reciprocity. *Plos ONE*, 16(8), e0255071.

[56] Wagner, R. L. (2019). Lowering consumers' price image without lowering their internal reference price: the role of pay-what-you-want pricing mechanism. *Journal of Revenue and Pricing Management*, 18(4), 332–341.

[57] Weisstein, F. L., Choi, P., & Andersen, P. (2019). The role of external reference price in pay-what-you-want pricing: An empirical investigation across product types. *Journal of Retailing and Consumer Services*, 50, 170–178.